\documentclass[12pt]{article}
\usepackage{epsfig}
\usepackage{caption2}

\title{Resource Sharing and Coevolution\\
in Evolving Cellular Automata}
\author{Justin Werfel\thanks{Department of Physics, Princeton University, 
Princeton, NJ 08544 (email: jkwerfel@princeton.edu)} 
\and Melanie Mitchell\thanks{Santa Fe Institute, 1399 Hyde Park Road, 
Santa Fe, NM 87501 (email: \{mm, chaos\}@santafe.edu)} \and James P. 
Crutchfield$^\dag$}

\setcaptionwidth{5.5in}

\begin{document}

\maketitle

\abstract{
Evolving one-dimensional cellular automata (CAs) with genetic
algorithms has provided insight into how improved performance on a
task requiring global coordination emerges when only local interactions
are possible. Two approaches that can affect the search efficiency of
the genetic algorithm are coevolution, in which a population of
{\it problems}---in our case, initial configurations of the CA
lattice---evolves along with the population of CAs; and resource
sharing, in which a greater proportion of a limited fitness resource
is assigned to those CAs which correctly solve problems that fewer
other CAs in the population can solve. Here we present evidence that,  in 
contrast to what has been suggested elsewhere,
the improvements observed when both techniques are used together depend
largely on resource sharing alone.

\section{Introduction}

Using evolutionary algorithms to design problem-solving strategies
often involves the use of {\em test cases} to estimate fitness, since the
space of possible strategies is typically too large to evaluate
exhaustively.
An important issue for improving statistical estimates of fitness in
such situations is how to sample test cases and then weight their
contribution to fitness estimates. This is particularly significant if
one wishes to avoid {\em premature convergence}, in which a mediocre 
solution strategy with no nearby fitter
variants takes over the population and prevents the emergence of
better solutions.

Techniques that have been proposed to ameliorate this difficulty include
{\em shared sampling}, in which test cases are chosen so as to be
unsolvable by as many of the strategies in the population as possible
\cite{Rosin&Belew95,Rosin&Belew97}; {\em competitive fitness functions},
in which a tournament-style selection scheme determines that one
strategy is fitter than another if the number of test cases solved by the
first, but not by the second, is greater than the number solved by the
second, but not the first \cite{Juille&Pollack97}; and
{\em resource-sharing fitness functions}, in which strategies receive a
higher fitness if they are able to solve test cases that are unsolvable
by a large fraction of other strategies
\cite{Juille&Pollack96,Juille&Pollack97,Rosin&Belew95,Rosin&Belew97}.

The motivation behind resource sharing is to promote diversity by
rewarding strategies that can solve test cases that few other
strategies are also able to solve. In this way strategies receive less
payoff for pursuing approaches that put them into ``niches'' already
heavily occupied. Instead they are encouraged to explore new
approaches, particularly those which allow solving test cases that the
population as a whole finds difficult. Presumably, the population ends
up more spread out over the space of possible strategies. In other
words, resource sharing is intended to preserve diversity, to prevent
mediocre solutions from taking over the population, and to make more
likely the emergence of exceptional new strategies through recombinations
of previously discovered strategies.

Another technique that has been proposed to improve the strategies
discovered by evolutionary search methods is that of coevolution, as
introduced by Hillis \cite{Hillis90}. Any particular static method for
generating test cases strongly affects the evolutionary course of
strategies. Moreover, there appears to be no single best method. If the
test cases are too easy, there is no strong pressure for
high-performance strategies to emerge; if the test cases are too hard,
then all low-performance strategies appear equally poor, reducing
fitness variance, and evolution cannot proceed.

In a coevolutionary scheme, a population of test cases is maintained
and evolves along with the evolving population of strategies. The
fitness of a strategy is then some function of the number of test cases
it is able to solve and the fitness of a test case is some inverse
function of the number of strategies that are able to solve it, often
with some penalty for being too difficult a test. The desired effect is
that the test-case population will evolve so as to present an
incrementally increasing but appropriate level of difficulty for the
evolving population that forces strategies to become successively more
capable of solving hard problems.

Past work \cite{Paredis97}, in accord with our own earlier
investigations, showed that a straightforward version of coevolution,
on its own, fails to produce high-performing strategies in a task in which 
cellular automata were evolved to perform a computation. The two
populations---candidate solutions and a set of test cases---enter
temporal oscillations in which each in turn performs well against the
other population. The individuals in both populations, however,
generally perform poorly against opponents chosen from outside the
populations.  Resource sharing has produced more promising results on 
other tasks
\cite{Juille&Pollack96,Juille&Pollack97,Rosin&Belew95,Rosin&Belew97}.

Combinations of different approaches for improving performance often
work better than each approach alone \cite{Rosin&Belew97}. In
particular, Juill\'e and Pollack
\cite{Juille&Pollack98a,Juille&Pollack98b} recently investigated a
combination of coevolution and resource sharing in evolving
cellular automata (CAs) to perform a density classification task
\cite{Crutchfield&Mitchell95a,CrutchfieldEtAl98,DasEtAl94a}, where the
CAs play the part of the strategies, and the initial configurations of
the CA lattice act as the test cases. Juill\'e and Pollack found that
using both approaches led to the production of significantly better CA
strategies than did the use of neither approach. They attributed this
success to the effectiveness of coevolution.

Since a somewhat different version of coevolution, acting alone, has
been shown not to produce effective strategies for this problem
\cite{Paredis97}, it seems natural to ask whether Juill\'e and Pollack's
success is due more to coevolution or to resource sharing, or to their
particular combination of the two.

\section{Methods and Results}

In Refs. \cite{Juille&Pollack98a} and \cite{Juille&Pollack98b},
Juill\'e and Pollack described results of
combining resource sharing and coevolution in the evolving cellular
automata framework of Crutchfield, Das, and Mitchell
\cite{Crutchfield&Mitchell95a,CrutchfieldEtAl98,DasEtAl94a}.  In that
framework, a genetic algorithm (GA) was used to evolve cellular automata
rule tables ({\em strategies}) to perform a density classification task.
The {\it fitness} of each strategy was a function of its
classification performance on a random sample of test cases: initial
configurations (ICs) of the CA lattice.  The ultimate success of the
GA was measured in terms of (1) the {\it performances} of the best
evolved strategies---their classification performance on larger, more
difficult sets of test cases than were used to calculate fitness
during evolution, and (2) the GAÕs {\em search efficiency}: the percentage
of runs on which high-performance strategies were evolved; see Ref.
\cite{CrutchfieldEtAl98} for details.

In Refs. \cite{Crutchfield&Mitchell95a,CrutchfieldEtAl98,DasEtAl94a},
we identified three classes of CA computational strategy evolved by
the GA, only one of which ({\em particle-based}) resulted in high
performance and generalized well to large lattice sizes. On 149-cell
lattices, these sophisticated strategies had performances ${\cal P}$
of 0.7 or greater.  We define {\em search efficiency} ${\cal E}_{\cal
P}$ as the percentage of runs on which strategies with performance
${\cal P}$ are evolved (i.e., at least one CA with performance $\geq
{\cal P}$ appears in the population).  In our original experiments,
${\cal E}_{0.7}$ was approximately 3\% . For reference, we note that
to date the best known CAs for density classification, evolved or
designed by hand, have performances above $0.8$ (on a scale of 0.0 to
1.0).

Juill\'e and Pollack showed that a particular combined form of resource
sharing and coevolution resulted in strategies with higher performance
(${\cal P} \approx 0.86$) \cite{Juille&Pollack98a,Juille&Pollack98b}
and a larger percentage of GA runs that produced high-performance
(${\cal P} \geq 0.7$) strategies; namely, search efficiencies above
30\% (H.  Juill\'e, personal communication).

As noted above, coevolution alone produced only low-performance
(${\cal P} << 0.7$) CAs \cite{Paredis97}. The results are
substantially worse than those of a GA alone---that is, a GA without
coevolution and without resource sharing. In particular, coevolution
does not produce high-fitness particle-based CAs and so results in a
search efficiency ${\cal E}_{0.7}$ of 0\%. Thus, we need not consider
this alternative GA further.

The experiments described here used GA and CA parameters, resource
sharing fitness functions, and a coevolution scheme identical to those
of Juill\'e and Pollack. The experiments were designed to probe their
effects in more detail than reported in Refs. \cite{Juille&Pollack98a}
and \cite{Juille&Pollack98b}.  

The populations of CAs and ICs each had 200 members. The CAs were
tested on 149-cell lattices.  We performed three experiments, each
consisting of 50 GA runs initiated with independent random number
seeds, where each run consisted of 1000 generations. The experiments
evaluated three search techniques: (1) the GA alone, with neither
resource sharing nor coevolution, with ICs drawn from a
density-uniform distribution (any IC density was equally probable)
(the same algorithm was used in Refs.
\cite{Crutchfield&Mitchell95a,CrutchfieldEtAl98,DasEtAl94a} but with
smaller values for population size and number of generations, and
subsequent lower search efficiencies); (2) the GA with resource
sharing only, with ICs drawn from a density-uniform distribution; and
(3) the GA with resource sharing and coevolution combined.  Under this
alternative, ICs were initially drawn from a denity-uniform
distribution and allowed to evolve thereafter.

For each experiment, we recorded the number of runs in which some
individual reached a performance level ${\cal P}$ or greater for four
values of ${\cal P}$: 0.65, 0.7, 0.75, and 0.8. From this we estimated
the mean generation $t_{\cal P}$ of first occurrence of threshold
${\cal P}$. The results are given in Table \ref{stats}. The standard
deviation $\sigma_{t_{\cal P}}$ of $t_{\cal P}$ across the 50 runs
of each alternative GA  is also reported there. In addition, the table
gives each alternative's observed search efficiencies.

\begin{table}[t]
\begin{center}
\begin{tabular}{|c|c|c|c|c|c|c|c|c|}
\hline
Search & Performance & Efficiency: & \multicolumn{2}{|c|}{Generations to 
$\cal P$} & \multicolumn{2}{|c|}{CAs: ${\cal P} > 0.8$} & 
\multicolumn{2}{|c|}{Diversity}\\ \cline{4-7}
Technique & Threshold $\cal P$ & \% Runs at $\cal P$ & $t_{\cal P}$ & 
$\sigma_{t_{\cal P}}$ & $\langle {\cal P} \rangle$ & ${\cal P}_{\rm best}$ & 
$s$ & $\sigma_s$\\
\hline\hline
Neither          & 0.65 & 100\% & 14 & 6 & & & & \\ \cline{2-5}
coevolution      & 0.70 & 29\% & 397 & 380 & 0.807 & 0.817 & -4.7 & 2.30\\ 
\cline{2-5}
nor resource     & 0.75 & 18\% & 231 & 363 & & & & \\ \cline{2-5}
sharing          & 0.80 & 6\%  & 488 & 442 & & & & \\
\hline\hline
Resource         & 0.65 & 100\% &  33 &  63 & & & & \\ \cline{2-5}
sharing          & 0.70 & 43\%  & 316 & 287 & 0.809 & 0.817 & 1.3 & 2.49\\ 
\cline{2-5}
only             & 0.75 & 37\%  & 276 & 221 & & & & \\ \cline{2-5}
                 & 0.80 & 10\%  & 479 & 127 & & & & \\
\hline\hline
Resource         & 0.65 & 100\% & 84 & 157 & & & & \\ \cline{2-5}
sharing and      & 0.70 & 47\% & 289 & 266 & 0.811 & 0.826 & 7.7 & 1.90\\ 
\cline{2-5}
coevolution      & 0.75 & 45\% & 295 & 249 & & & & \\ \cline{2-5}
                 & 0.80 & 27\% & 438 & 235 & & & & \\
\hline
\end{tabular}
\end{center}
\caption{Statistics for the evolutionary emergence of CAs with performance
$\cal P$ exceeding various thresholds. The percentage of the $50$ runs
reaching threshold $\cal P$ is given as an estimate of each alternative
GA's search efficiency ${\cal E}_{\cal P}$. $t_{\cal P}$ is the mean number of 
generations to
first occurrence of performance ${\cal P}$ estimated across $50$ GA runs.
$\sigma_{t_{\cal P}}$ is the standard deviation in $t_{\cal P}$ estimated
across the runs. $s$ is the rate of change in population diversity
$\langle d \rangle$ (quoted in bits per $1000$ generations) and $\sigma_s$
its standard deviation estimated in the least-squares fits of Figs.
\ref{hams}(a), \ref{hams}(b), and \ref{hams}(c).
}
\label{stats}
\end{table}

The results for experiments involving resource sharing agree, within
statistical uncertainty, with results found by Juill\'e for the
percentage of runs in which a CA exceeding these thresholds occurs and
for the mean first generation of such occurrences (H. Juill\'e, 
personal communication). 

Runs with resource sharing alone produce CAs with higher performance
more consistently across GA runs (e.g., ${\cal E}_{0.7} = 43\%$ rather
than 29\%). Moreover, runs with resource sharing tend to take longer
($t_{\cal P} \approx 33$, rather than $14$) to find moderate-performance
CAs (${\cal P} \geq 0.65$). They also vary more
($\sigma_{t_{\cal P}} \approx 63$, rather than $6$) in how long they
take to do so, than runs without resource sharing. Comparing runs using
both resource sharing and coevolution to those using resource sharing
alone, the addition of coevolution appears to heighten the effects of
resource sharing. Runs using both techniques take longer
($t_{\cal P} \approx 84$, rather than $33$) to find moderate-performance
CAs, vary more in how long it takes them to do so
($\sigma_{t_{\cal P}} \approx 157$ rather than $63$), and find
high-performance CAs  more frequently (e.g., ${\cal E}_{0.7} =  47\%$ rather 
than 43\% and ${\cal E}_{0.8} = 27\%$ rather than 10\% ) than do runs with 
resource sharing alone.

As an aside, note that the large variances $\sigma_{t_{\cal P}}$ in mean
time to a given fitness threshold are typical of and to be expected in
evolutionary search algorithms. The nature of such fluctuations is
discussed in Ref. \cite{Nimw98b}. What is notable is that the GA using
resource sharing appears to have roughly half the variation seen in
the GA alone. The addition of coevolution to resource sharing
appears to have little (beneficial) effect in reducing the variations
for the appearance times of high-performance CAs. In fact, in reaching
the highest-performing CAs, the addition of coevolution roughly doubles
the variance in $t_{\cal P}$.

What about the ability of one or the other alternative GA to reach
high-performance CAs? Except for the very last performance threshold
(${\cal P} > 0.8$), the differences between resource sharing with
coevolution and resource sharing alone are less pronounced than the
corresponding differences between resource sharing alone and neither
approach. Given the poor evolved-CA performance and 0\% search efficiency 
${\cal E}_{0.7}$
of coevolution-only GAs, it thus appears that resource sharing must be
used if the addition of coevolution is to improve the GA's performance.
While that latter improvement is helpful, we conclude that resource
sharing, rather than coevolution, is the key technique leading to
improvements in the performance of CAs evolved by the GA and in the
frequency of their discovery across GA runs.

\section{The operation of resource sharing}

We may ask further whether the effectiveness of resource sharing is
actually due, as was intended in its design, to a preservation of
diversity in the respective populations, or whether its success results
from some other mechanism entirely.

One rough measure of diversity in a population is the average pairwise
genetic Hamming distance $\langle d \rangle$. The Hamming distance
$d$ between two CAs is simply the number of bits by which the genetic
specification of their update rules differ. CAs with different strategies
are likely to differ in more bits and thus to be separated by a greater
Hamming distance than CAs with similar strategies. When averaged over the
population, $\langle d \rangle$ is greater if a population is more
strategically diverse overall and its members are more spread out across
the genotype space. Concerned about possible long-tailed distributions
governing $d$, we estimated median, in addition to average, pairwise
Hamming distances. There was no qualitative change to the results.
Moreover, the median distance never differed from the average by more
than a single bit after the first few generations. For these reasons,
we report here only average Hamming distances.

To give a sense of the scale of Hamming distances here, we recall
several facts about the type of CA evolved. The state of an individual
cell in one of our CAs is determined by its own state and the states
of its six nearest neighbors at the previous time step. There are then
$7$ bits in the input to a CA's rule table. Since the cell states are
binary, the rule table is specified by $2^7 = 128$ output bits. Thus,
$0 \leq \langle d \rangle \leq 128$.

\begin{figure}[t]
\begin{center}
\epsfig{file=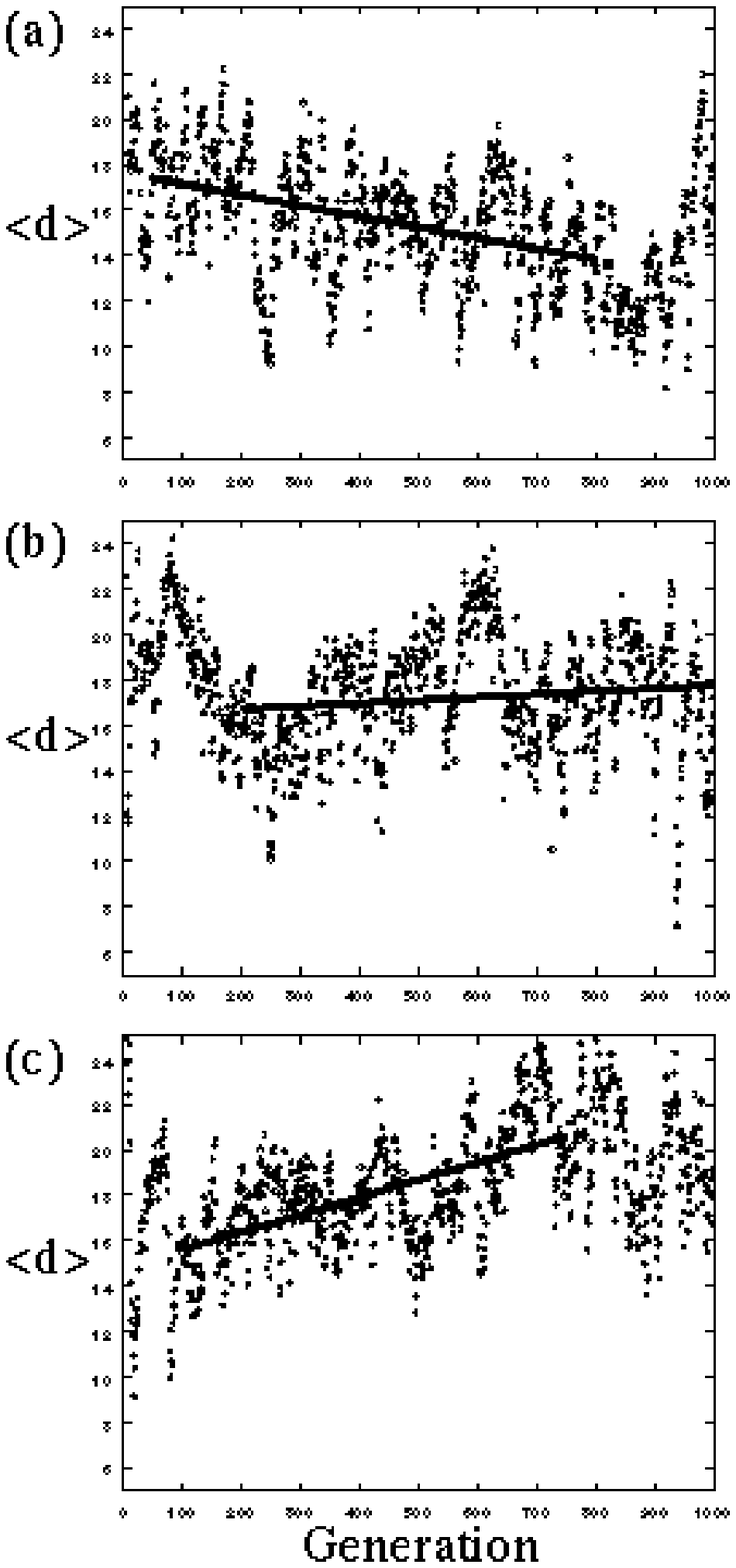}
\caption{Average pairwise Hamming distance $\langle d \rangle$ over
  time for single GA runs with (a) neither resource sharing nor
  coevolution, (b) resource sharing alone, (c) resource sharing and
  coevolution. $\langle d \rangle$ is large ($\approx 64$) during the 
  initial generations, and so these data points do not appear on the scales
  plotted. The straight lines show the trends in population diversity.
  They are least-squares fits over stationary fitness epochs in the
  population dynamics. The estimated slopes $s$ and their standard 
deviations
  $\sigma_s$ are quoted in Table \ref{stats}. The runs shown here are
  typical of those that evolved high-fitness particle-based CAs under
  each alternative GA.}
\label{hams}
\end{center}
\end{figure}

Figure \ref{hams} shows $\langle d \rangle$ at each generation for
runs---typical of those that evolved high-fitness particle-based
CAs---with (a) neither resource sharing nor coevolution, (b) resource
sharing alone, and (c) both techniques. In all cases,
$\langle d \rangle$ starts out very large
($\langle d \rangle \approx 128/2 =64 $ bits), since the initial CA
population was randomly initialized. $\langle d \rangle$ quickly
decreases over a few generations, as the fittest CAs and their
descendants take over the population, which settles down
to CAs with similar strategies.

Beyond the transient phase, over each run $\langle d \rangle$
fluctuates about 10\% to 20\% as evolution progresses. Nonetheless, as
the plots show, each run does follow an overall trend in population
diversity. We measured these trends using a least-squares fit to
estimate the average rate $s$ of change in population diversity.  We
also estimated the standard deviation $\sigma_s$ of the fit. Both
estimates for each run are reported in Table \ref{stats}. Since such
trends are interrupted as the GA discovers progressively more
effective CAs, the fits were made only over a stationary "epoch" (a
period in which average population fitness remains roughly constant).
For Fig. \ref{hams}(a), the fit is made for an epoch lasting from
generation $50$ to generation $800$; in Fig. \ref{hams}(b), from
generation $200$ to $1000$; and in Fig. \ref{hams}(c), from $100$ to
$750$.

In runs with neither resource sharing nor coevolution,
$\langle d \rangle$ decreases slowly over time, with temporary
increases each time a new, more effective type of strategy is
discovered. In Figure \ref{hams}(a), for example, $\langle d \rangle$
declines slowly over nearly $800$ generations, from about $18$ bits to a
minimum of approximately $12$ bits. At that point, a new strategy
appears around generation $900$ and $\langle d \rangle$ increases again.
The estimated trend shows a negative slope and one concludes that with
this type of GA population diversity steadily decreases in an epoch.

In contrast, when the GA with resource sharing is used, as illustrated
by the run in Fig. \ref{hams}(b), $\langle d \rangle$ remains roughly
constant around $17.5$ bits. There also appears to be wider fluctuation
in the population diversity about this trend than in the alternative GAs;
see $\sigma_s$ in Table \ref{stats}.

When coevolution is added to resource sharing, as shown by the run in
Fig. \ref{hams}(c), $\langle d \rangle$ increases over time, after the
population initially settles down. $\langle d \rangle$ typically goes
from about $15$ bits to about $21$ bits over the course of 700
generations.

It would appear then that resource sharing maintains diversity, as it
was intended to do.  Its use prevents the slow decrease in total Hamming
distance that otherwise occurs as the population converges on a
narrower range of strategies. In other words, it maintains a wider variation in 
the space of CAs. The addition of coevolution appears to enhance
the effect of resource sharing, when the latter is also used: the
total Hamming distance increases and reaches noticeably higher values
over a similar number of generations.

\section{Further Work}

If coevolution does, in fact, augment the operation of resource sharing, as
suggested by the above statistical analysis, it remains unclear exactly
how it acts to do so. A more detailed analysis of how the GA with resource
sharing is affected by the addition of coevolution, and why coevolution is
useful when it accompanies resource sharing but not on its own, will be
necessary for a fuller understanding of the trade-offs between these
alternative evolutionary search techniques. This is motivated, of course,
by the marked and very useful improvements shown by the augmented GAs
over the GA alone.

Underlying these overall concerns and determining, in large measure,
the statistical analysis just reported is a complicated problem in
nonlinear population dynamics. High-performance CAs evolve via a series
of epochs of stasis punctuated by sudden innovations
\cite{Crutchfield&Mitchell95a,CrutchfieldEtAl98,DasEtAl94a},
whether resource sharing, resource sharing and coevolution, or neither
are employed. The dynamics of epochal evolution has recently been
mathematically analyzed in some detail; see Refs.
\cite{Nimw98b,Nimw97b,Nimw98a}.  It would be useful, therefore, to bring
the current investigations together with this mathematical analysis to
understand why epochal evolution with resource sharing results in
higher variance in the time it takes moderate- and high-performance
CAs to emerge, why higher-performance CAs appear more frequently with
resource sharing, and how it is that coevolution increases these
effects. Here we have begun to understand more systematically
how resource sharing and coevolution affect the evolutionary process,
but not yet their underlying mechanisms.

\section{Acknowledgments}
We thank Rajarshi Das and Wim Hordijk for their help in performing
these experiments and Hugues Juill\'e for helpful discussions. We
thank Wim Hordijk for comments on the manuscript. This
work was supported by the Santa Fe Institute,  by the National
Science Foundation
grants PHY-9531317 (Research Experiences for Undergraduates) and  IRI-
9705830, and by the Keck Foundation grant 98-1677.

%{\bf BIBLIOGRAPHY: Refs [5,6,7,8,10,11] need page ranges.}

\bibliography{rsc}
\bibliographystyle{plain}

\end{document}